\documentclass[prd,showpacs]{revtex4}

\usepackage{amsmath}
\usepackage{amsthm}
\usepackage{amssymb}
\usepackage{amscd}
\usepackage[british]{babel}
\usepackage{graphicx}
\usepackage{psfrag}
\usepackage{epsfig}
\usepackage{rotating}
\usepackage{times}

\begin{document}

\title{On the fate of the phantom dark energy universe in semiclassical gravity II: \\ Scalar phantom fields}

\author{Jaume de Haro$^{1,}$\footnote{E-mail: jaime.haro@upc.edu},
Jaume Amoros$^{1,}$\footnote{E-mail: jaume.amoros@upc.edu} and
Emilio Elizalde$^{2,}$\footnote{E-mail: elizalde@ieec.uab.es,
elizalde@math.mit.edu}}

\affiliation{$^1$Departament de Matem\`atica Aplicada I, Universitat
Polit\`ecnica de Catalunya, Diagonal 647, 08028 Barcelona, Spain \\
$^2$Instituto de Ciencias del Espacio (CSIC) \& Institut
d'Estudis Espacials de Catalunya (IEEC/CSIC)\\ Campus UAB, Facultat
de Ci\`encies, Torre C5-Parell-2a planta, 08193 Bellaterra
(Barcelona) Spain}

\pagestyle{myheadings}

\theoremstyle{plain}
\newtheorem{lemma}{Lemma}[section]
\newtheorem{theorem}{Theorem}[section]
\newtheorem{proposition}{Proposition}[section]
\newtheorem{corollary}{Corollary}[section]
\newtheorem{remark}{Remark}[section]
\newtheorem{definition}{Definition}[section]
\newtheorem{example}{Example}[section]

\newcommand{\boxend}{\flushright{$\Box$}}
\newenvironment{dem}[1]{\begin{trivlist} \item {\bf Proof #1\/:}}
            {\boxend \end{trivlist}}

\newcommand{\N}{{\mathbb N}}               
\newcommand{\Z}{{\mathbb Z}}               
\newcommand{\Q}{{\mathbb Q}}               
\newcommand{\R}{{\mathbb R}}               
\newcommand{\C}{{\mathbb C}}               
\renewcommand{\S}{{\mathbb S}}             
\newcommand{\T}{{\mathbb T}}               
\newcommand{\D}{{\mathbb D}}               

\newcommand{\half}{\frac{1}{2}}
\renewcommand{\Re}{\mbox{\rm Re}}
\renewcommand{\Im}{\mbox{\rm Im}}
\newcommand{\sprod}[2]{\left\langle#1,#2\right\rangle}
\newcommand{\inner}[2]{\left\langle#1,#2\right\rangle_2}

\newcommand{\e}{\epsilon}
\newcommand{\w}{\omega}
\newcommand{\f}{\frac}

\newcommand{\ep}{\varepsilon}
\newcommand{\al}{\alpha}
\newcommand{\h}{\hbar}
\renewcommand{\tilde}{\widetilde}

\thispagestyle{empty}

\begin{abstract}

\vspace{1cm}
Quantum corrections coming from massless fields conformally coupled
with gravity are studied, in order to see if they can lead to 
avoidance of the annoying Big Rip singularity which shows up in a flat
Friedmann-Robertson-Walker universe filled with dark energy
and modeled by a scalar phantom field.
The dynamics of the model are discussed for all values of the two
parameters, named $\alpha>0$ and $\beta<0$, corresponding to
the regularization process. The new  results are compared with the
ones obtained in \cite{hae11} previously, where dark energy was modeled 
by means of a phantom fluid with equation of state
 $P=\omega\rho$, with $\omega<-1$.
\end{abstract}


\maketitle

\vspace{-6.5mm}

\hspace*{10mm}  {\footnotesize Keywords: Dark energy, future
singularities, semiclassical gravity}


\section{Introduction}
Recent observations of distant type Ia supernovae, baryonic acoustic oscillations (BAO), anisotropies of the cosmic microwave background radiation (CMB), and some other,
confirm that our universe expands in an accelerated way \cite{p99,r99}. In fact,
it seems that it is at present in a dark energy phase \cite{k11}. A proposal to explain this situation is to assume that the energy density of our universe is
dominated by a phantom scalar field, that is, a model where the
energy density  and pressure are $\rho=-\frac{1}{2}\dot{\phi}^2+V(\phi)$ and $P=-\frac{1}{2}\dot{\phi}^2-V(\phi)$, being
$\phi$ the scalar phantom field.
In this case, future singularities are bound to appear in a finite time \cite{s07}. These singularities are undoubtedly there in the classical situation when no
quantum effects are taken into account, but it seems feasible that, near the singularities, where the curvature has very high values, quantum effects could have the power to drastically modify the behavior of the universe, yielding a milder singularity or maybe even a non-singular model.

In this paper we extend the study carried out in \cite{hae11} to a dark energy universe 
modeled by a scalar phantom field. In fact we will consider exponential potentials which give rise to a Big Rip singulary, and introduce quantum corrections in order to avoid these late time singularities. Specifically, we shall consider the quantum
effects due to massless, conformally coupled fields. This is a special, workable case where the
quantum vacuum stress tensor---which depends on two regularization
parameters, here called $\alpha>0$ and $\beta<0$---and the
semiclassical Friedmann equation, can be both calculated explicitly.

We will show, analytically and numerically, that quantum effects drastically modify the Big Rip singularity, rendering it of type III  or turning it into a singularity
in the contracting phase. In the first case (a type III singularity) the Hubble parameter does not diverge, but the energy density does tend towards
infinity. In the other case (a singularity in the contracting phase) the Hubble parameter becomes finite and negative, and the energy density diverges towards
minus infinity. What is important to note is that, in both cases, the Hubble parameter remains finite.

The paper is organized as follows. In the next Section, using the mathematical theory of dynamical systems,  we study some phantom
fields driving the universe to a Big Rip singularity. In Sect.~III we introduce the quantum corrections due to a massless conformally coupled field,
and we perform an analytic study of the semiclassical Friedmann equations. In Sect.~IV a numerical analysis is carried out to check to good approximation the analytic results obtained in the previous Section.
In Sect.~V we analyze the problem in the context of loop quantum cosmology, where it has been stated that quantum corrections do completely avoid the
 Big Rip singularity.  We will see  that the way to obtain these conclusion is in doubt, because they have been got in some places from
an incorrectly modified Friedmann equation.
In last Section we compare the results obtained for a phantom field with those that were derived for a phantom fluid model.
 The units to be used in the paper are: $c=\hbar=M_p=1$, where $M_p$ is the reduced Planck mass.

\section{Dark energy modeled by a phantom field}

A phantom field $\phi$ is modeled by an energy density of the kind $\rho=-\frac{1}{2}\dot{\phi}^2+V(\phi)$ and  a pressure
$P=-\frac{1}{2}\dot{\phi}^2-V(\phi)$. For this field the
Friedmann and conservation equations are:
\begin{eqnarray}\label{a1}\left\{\begin{array}{ccc}
H^2&=& \frac{1}{3}\left(-\frac{1}{2}\dot{\phi}^2+V(\phi)\right)\\
&&\\
0&=&-\ddot{\phi}-3H\dot{\phi}+\frac{d V}{d{\phi}}\end{array}\right.
\end{eqnarray}
From this system one deduces that $\dot{\rho}=3H\dot{\phi}^2$, and thus $\dot{H}=\frac{1}{2}\dot{\phi}^2>0$, which means that $\frac{\ddot{a}}{a}=\dot{H}+H^2>0$, that is, in this model the universe is expanding in an accelerating way.

To analyze the dynamics of the system we start considering a power-law potential, i.e., $V(\phi)=\lambda \phi^{2n}$ with $\lambda>0$.
The field equation can be written as follows
\begin{eqnarray}\label{a2}
 \frac{d}{dt}\left(\frac{1}{2}\dot{\phi}^2+\tilde V(\phi)\right)=-3H\dot{\phi}^2,
\end{eqnarray}
where $\tilde V(\phi)=-V(\phi)$. This is a dissipative system, and the slow-roll conditions $\left(V'/V\right)^2\ll 1$
and $\left|V''/V\right|\ll 1$ are satisfied when $|\phi|\gg n$. Then, due to the attractor nature of the slow-roll regime,  at
late time, the solutions have the same behavior as the slow-roll solution, which satisfies the system
\begin{eqnarray}\label{a3}\left\{\begin{array}{ccc}
H^2&=& \frac{1}{3}V(\phi)\\
&&\\
0&=&-3H\dot{\phi}+\frac{d V}{d{\phi}}\end{array}\right.
\end{eqnarray}
Since the dynamics of the system decouples for $\phi>0$ and $\phi<0$, we only consider the domain $\phi>0$, where the field obeys
the equation $\dot{\phi}_{sr}=2n\sqrt{\frac{\lambda}{3}}\phi_{sr}^{n-1}$, which solution is
\begin{eqnarray}\label{a4}\left\{\begin{array}{ccc}
 \phi_{sr}(t)=\left[2n(n-2)\sqrt{\frac{\lambda}{3}}(t_s-t)\right]^{\frac{1}{2-n}}& for  & n>2\\
\phi_{sr}(t)=\phi_{sr}(t_0)e^{4\sqrt{\frac{\lambda}{3}}(t-t_0)}& for  & n=2\\
\phi_{sr}(t)=\phi_{sr}(t_0)+2\sqrt{\frac{\lambda}{3}}(t-t_0)& for  & n=1.
\end{array}\right.
\end{eqnarray}
Evaluating $H_{sr}(t)=\sqrt{\frac{\lambda}{3}}\phi_{sr}^{n}(t)$, one can see that the Big Rip singularity appears for $n>2$.

The following remark is in order. To prove, in a rigorous way, the attractor nature of the slow roll solution, we may use the variables
$x\equiv \frac{\phi}{\sqrt{6}}$ and $y\equiv \frac{\dot{\phi}}{\sqrt{6}H}$ (\cite{m05}). Then, the dynamical equation is
\begin{eqnarray}\label{a5}
\frac{dy}{dx}=-3(1+y^2)\left[1-\frac{V'}{\sqrt{6}yV}\right]=-3(1+y^2)\left[1-\frac{2n}{{6}yx}\right].
\end{eqnarray}
The slow roll solution is the curve $\frac{dy}{dx}=0$, i.e., $y=\frac{2n}{{6}x}$, and it is easy to verify that, for large values of $x$, this is
the leading term of the solution. Then, since $\frac{dy}{dx}<0$ above this curve, and $\frac{dy}{dx}>0$ below it, this definitely proves that the slow
roll solution is an attractor at late times.

 As a different specific example, we choose  $V(\phi)=V_0e^{-2\phi/\phi_0}$ (being $V_0$ and $\phi_0$ two constant parameters). Then,  with the change of function
$\phi=\phi_0\ln(\psi)$ (now $\psi$ belongs in the domain $(0,\infty)$), the system becomes
\begin{eqnarray}\label{a6}\left\{\begin{array}{ccc}
H^2&=& \frac{1}{3\psi^2}\left(-\frac{\phi_0^2}{2}{\dot{\psi}^2}+V_0\right),\\
&&\\
0&=&-\phi_0(\ddot{\psi}\psi-\dot{\psi}^2)-3H\phi_0\dot{\psi}\psi-2\frac{V_0}{\phi_0}.\end{array}\right.
\end{eqnarray}
In the expanding phase $(H>0)$, this system can be written as follows:
\begin{eqnarray}\label{a7}\phi_0\ddot{\psi}\psi
+\frac{2}{\phi_0}\left(-\frac{\phi_0^2}{2}{\dot{\psi}^2}+V_0 \right)+\sqrt{3}\phi_0\dot{\psi}\sqrt{-\frac{\phi_0^2}{2}{\dot{\psi}^2}+V_0}=0,
\end{eqnarray}
and dividing this equation by $\dot{\phi}$---i.e., using the variable $\phi$ as a time---one obtains the equation
\begin{eqnarray}\label{a8}
\frac{d\dot{\psi}}{d\phi}=
-\frac{2}{\phi_0^3\dot{\psi}}\left(-\frac{\phi_0^2}{2}{\dot{\psi}^2}+V_0 \right)-\frac{\sqrt{3}}{\phi_0}
\sqrt{-\frac{\phi_0^2}{2}{\dot{\psi}^2}+V_0}\equiv F(\dot{\psi}).
\end{eqnarray}
This is an autonomous first order differential equation, therefore, it can be completely studied just through the sign of the function $F$. From the Friedmann equation, one can see
that the domain of $F$ is the interval $[-\frac{\sqrt{2V_0}}{\phi_0},\frac{\sqrt{2V_0}}{\phi_0}]$.
The zeros of $F$ are the points $\pm\frac{\sqrt{2V_0}}{\phi_0}$
and $-\frac{\sqrt{2V_0}}{\phi_0}\frac{1}{\sqrt{1+\frac{3}{2}\phi_0^2}}$.
$F$ has a vertical asymptotic at zero. Finally, $F$ is positive in the interval
$(-\frac{\sqrt{2V_0}}{\phi_0}\frac{1}{\sqrt{1+\frac{3}{2}\phi_0^2}},0)$ and negative in
$(-\frac{\sqrt{2V_0}}{\phi_0},-\frac{\sqrt{2V_0}}{\phi_0}\frac{1}{\sqrt{1+\frac{3}{2}\phi_0^2}})\cup (0,\frac{\sqrt{2V_0}}{\phi_0})$.

This all means that, using $\phi$ as time, the critical points $\frac{\sqrt{2V_0}}{\phi_0}$
and $-\frac{\sqrt{2V_0}}{\phi_0}\frac{1}{\sqrt{1+\frac{3}{2}\phi_0^2}}$ are repellers and
the critical points $0$ and $-\frac{\sqrt{2V_0}}{\phi_0}$ are attractors (see Pict.~$1$).
However, in the domain $\dot{\psi}<0$, when the time $\phi$ increases, the cosmic time $t$ decreases, and vice versa, what means that, in terms of the cosmic time, the critical point $ -\frac{\sqrt{2V_0}}{\phi_0}\frac{1}{\sqrt{1+\frac{3}{2}\phi_0^2}}$
is a global attractor while the other two critical points $\pm\frac{\sqrt{2V_0}}{\phi_0}$ are repellers (see Pict.~$2$).



In terms of the field $\phi$, the critical points obtained above are:
\begin{eqnarray}\label{a9}
\dot{\psi}=-\frac{\sqrt{2V_0}}{\phi_0}\frac{1}{\sqrt{1+\frac{3}{2}\phi_0^2}}\Longleftrightarrow
\phi(t)=\phi_0\ln\left(\frac{t_s-t}{\frac{\phi_0}{\sqrt{2V_0}}\sqrt{1+\frac{3}{2}\phi_0^2}}\right);\quad  H(t)=\frac{\phi_0^2/2}{t_s-t},\nonumber\\
\dot{\psi}=\pm\frac{\sqrt{2V_0}}{\phi_0}\Longleftrightarrow
\phi(t)=\phi_0\ln\left(\frac{t_s-t}{\mp\frac{\phi_0}{\sqrt{2V_0}}}\right);\quad  H(t)=0,
\end{eqnarray}
where $t_s$ is an arbitrary constant.
Then, since $-\frac{\sqrt{2V_0}}{\phi_0}\frac{1}{\sqrt{1+\frac{3}{2}\phi_0^2}}$ is a global attractor, it follows that, 
except for the solutions $\dot{\psi}=\pm\frac{\sqrt{2V_0}}{\phi_0}$, all the other have a Big Rip singularity.

\begin{eqnarray*}
\begin{picture}(74,44)\thicklines
\put(-40,-10){\line(1,0){240}}
\put(-80,-10){\line(1,0){100}}
\put(30,-10){\vector(1,0){2}}
\put(-60,-10){\vector(-1,0){2}}
\put(120,-10){\vector(-1,0){2}}
\put(200,-10){\circle*{3}}
\put(-80,-10){\circle*{3}}
\put(-20,-10){\circle*{3}}
\put(50,-10){\circle*{3}}
\put(200,-14){\makebox(0,0)[t]{$\frac{\sqrt{2V_0}}{\phi_0}$}}
\put(-80,-14){\makebox(0,0)[t]{$-\frac{\sqrt{2V_0}}{\phi_0}$}}
\put(-20,-14){\makebox(0,0)[t]{$-\frac{\sqrt{2V_0}}{\phi_0}\frac{1}{\sqrt{1+\frac{3}{2}\phi_0^2}}$}}
\put(50,-14){\makebox(0,0)[t]{$0$}}
\put(50,20){\makebox(0,0)[t]{ time $\phi$}}
\put(50,-35){\makebox(0,0)[t]{Picture $1$}}
\end{picture}
\end{eqnarray*}

\vspace{2cm}

\begin{eqnarray*}
\begin{picture}(74,44)\thicklines
\put(-40,-10){\line(1,0){240}}
\put(-80,-10){\line(1,0){100}}
\put(30,-10){\vector(-1,0){2}}
\put(-60,-10){\vector(1,0){2}}
\put(120,-10){\vector(-1,0){2}}
\put(200,-10){\circle*{3}}
\put(-80,-10){\circle*{3}}
\put(-20,-10){\circle*{3}}
\put(50,-10){\circle*{3}}
\put(200,-14){\makebox(0,0)[t]{$\frac{\sqrt{2V_0}}{\phi_0}$}}
\put(-80,-14){\makebox(0,0)[t]{$-\frac{\sqrt{2V_0}}{\phi_0}$}}
\put(-20,-14){\makebox(0,0)[t]{$-\frac{\sqrt{2V_0}}{\phi_0}\frac{1}{\sqrt{1+\frac{3}{2}\phi_0^2}}$}}
\put(50,-14){\makebox(0,0)[t]{$0$}}
\put(50,20){\makebox(0,0)[t]{ time $t$}}
\put(50,-35){\makebox(0,0)[t]{Picture $2$}}
\end{picture}
\end{eqnarray*}

\vspace{2cm}


\begin{figure}[h]
\begin{center}
\includegraphics[scale=0.50]{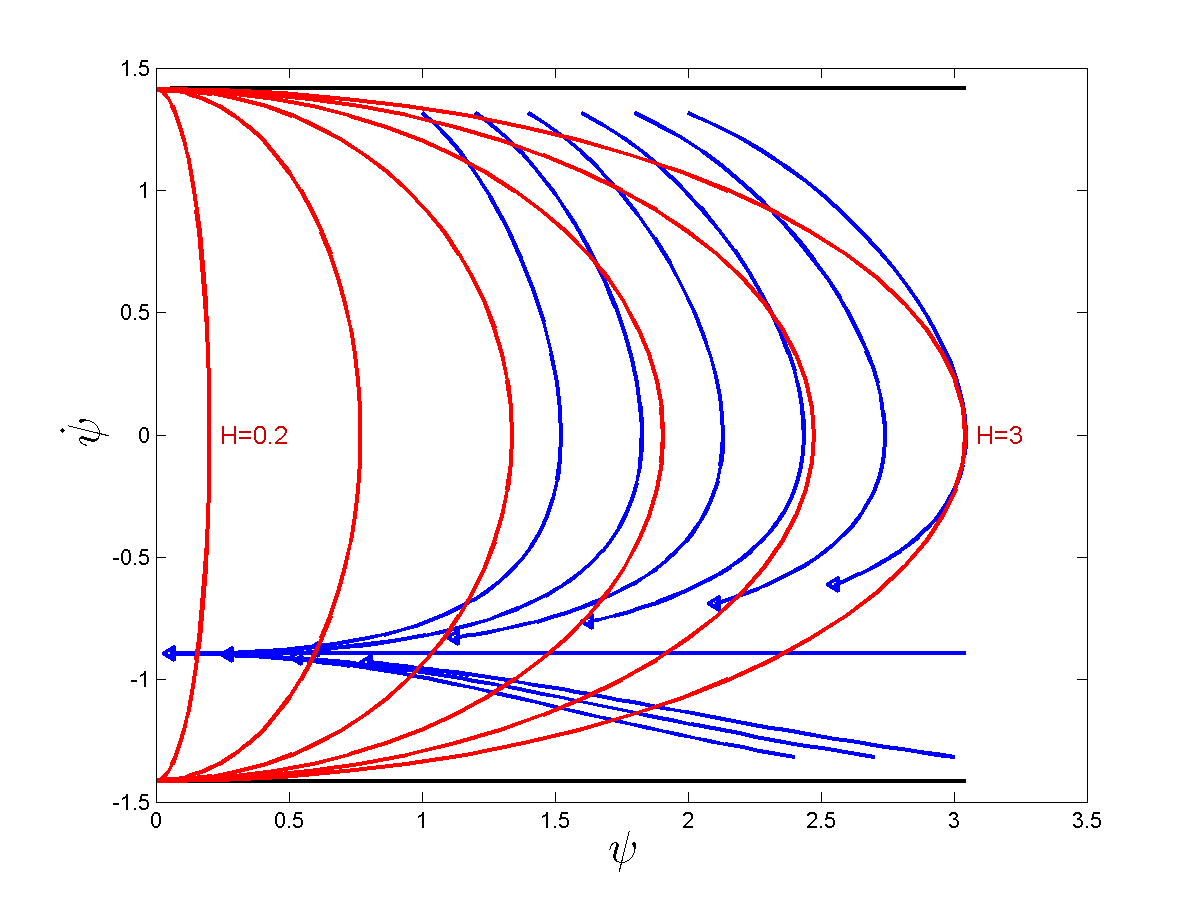}
\end{center}

\vskip-6mm
\caption{{Phase space portrait for $\phi_0=V_0=1$. Blue curves are the orbits of the system and the red ones are the level curves of $H$. The
line $\dot{\psi}=-\frac{2}{\sqrt{5}}$ is the global attractor orbit.}}
\end{figure}

\section{Quantum corrections}
In this section we will study in detail what is the change in the dynamics in the model $V(\phi)=V_0e^{-2\phi/\phi_0}$ when one takes into account quantum
effects. It is well-known that
for a massless, conformally coupled field, the anomalous trace is given by \cite{hae11,not05}
\begin{eqnarray}\label{a10}
 T_{vac}= \alpha\Box  R-\frac{\beta}{2}G,
\end{eqnarray}
with  $R=6\left(\dot{H}+2H^2\right)$  the scalar curvature and
$G=24H^2\left(\dot{H}+H^2\right)$ the
Gauss-Bonnet curvature invariant.

The coefficients, $\alpha$ and $\beta$, coming from dimensional regularization are
\cite{fhh79}
\begin{eqnarray}\label{a11}
\alpha=\frac{1}{2880\pi^2}(N_0+6N_{1/2}+12N_1)>0,\nonumber\\
\beta=\frac{-1}{2880\pi^2}(N_0+11N_{1/2}+62N_1)<0,
\end{eqnarray}
being $N_0$ the number of scalar fields, $N_{1/2}$ the number
of four-component neutrinos, and $N_1$ the number of electromagnetic
fields, respectively.

In terms of the Hubble parameter,
Eq.~(\ref{a10}) is \cite{d77}
\begin{eqnarray}\label{a12}
T_{vac}=6\alpha(\dddot{H}+12H^2\dot{H}+7H\ddot{H}+4\dot{H}^2)-12\beta
(H^4+H^2\dot{H}).
\end{eqnarray}
With the trace anomaly being $T_{vac}=\rho_{vac}-3P_{vac}$ and,
inserting (\ref{a12}) into the conservation equation,
$\dot{\rho}_{vac}+3H(\rho_{vac}+P_{vac})=0$, the
modified  energy density reads
\begin{eqnarray}\label{a13}
\rho_{vac}=6\alpha(3H^2\dot{H}+H\ddot{H}-\frac{1}{2}\dot{H}^2)-3\beta
H^4,
\end{eqnarray}
and the semiclassical Friedmann equation becomes
\begin{eqnarray}\label{a14}
 H^2=\frac{\rho+\rho_{vac}}{3},
\end{eqnarray}
with $\rho=-\frac{1}{2}\dot{\phi}^2+V_0e^{-2\phi/\phi_0}$.

Using the dimensionless variables
$\bar{t}=H_+t$, $\bar{H}=H/H_+$, $\bar{Y}=\dot{H}/H_+^2$, $\bar{\psi}=\psi$   and
$\bar{\rho}=\frac{\rho}{3H_+^2}$, with $H_+=\sqrt{-1/\beta}$, the
semiclassical Friedmann equation and the conservation equation can be
written as an autonomous system:
\begin{eqnarray}\label{a15}\left\{\begin{array}{ccl}
{\bar{H}}'&=&\bar{Y},\\
{\bar{Y}}'&=& \frac{1}{2\alpha \bar{H}}\left(-\beta{\bar{H}^2}+\beta\bar{\rho}
-6\alpha \bar{H}^2\bar{Y}+\alpha \bar{Y}^2+\beta \bar{H}^4\right),\\
{\bar{\psi}}'&=&\bar{\varphi},\\
\bar{\varphi}'&=&\frac{\bar{\varphi}^2}{\bar{\psi}}-3\bar{H}\bar{\varphi}-
2\frac{\bar{V_0}}{\bar{\phi_0}^2\bar{\psi}},\end{array}\right.\end{eqnarray}
where $'$ denotes derivative with respect to the time $\bar{t}$, and we have defined the new parameters $\bar{V}_0=V_0/H_+^2$ and $\bar{\phi}_0=\phi_0$.

What we see at first sight from this system is that it does not have any critical point. It is also easy to show that the energy density $\bar{\rho}$ evolves in accordance with the equation ${\bar{\rho}}'=\bar{H}\bar{\phi}_0^2\frac{\bar{\varphi}^2}{\bar{\psi}^2}$, which means that
the energy density increases in the expanding phase, and decreases in the contracting one.

Now, we look for singular solutions of the system with the following behavior near to the singularity \cite{no04a,eno04}
\begin{eqnarray}\label{a16}
 \bar{\psi}(\bar{t})=A(\bar{t}_s-\bar{t})+B(\bar{t}_s-\bar{t})^2+{\mathcal O}\left((\bar{t}_s-\bar{t})^3\right);\quad
\bar{H}(\bar{t})=\bar{H}_0+\delta\bar{H}(\bar{t}),
\end{eqnarray}
where $A$, $B$ and $\bar{H}_0$ are some constants. Inserting these functions in the conservation equation, one obtains
\begin{eqnarray}\label{a17}
 A=\frac{\sqrt{2\bar{V}_0}}{\bar{\phi}_0};\quad  B=-\frac{3}{2}\bar{H}_0 A,
\end{eqnarray}
and
inserting them in the semiclassical Friedmann equation and retaining the leading terms, one gets
\begin{eqnarray}\label{a18}
 \delta\bar{H}''(\bar{t})=\frac{\beta}{2\alpha}\frac{\bar{\phi}_0^2}{\bar{t}-\bar{t}_s}\Rightarrow
\delta\bar{H}(\bar{t})=\frac{\beta}{2\alpha}\bar{\phi}_0^2(\bar{t}_s-\bar{t})\ln\left((\bar{t}_s-\bar{t})/T\right),
\end{eqnarray}
where $T$ is an integration constant.

What we observe here is that, when we introduce quantum corrections, the Big Rip singularity, for $\bar{H}_0>0$, is transformed into a type III singularity, because
as $\bar{t}\rightarrow \bar{t}_s$ one has  ${H}\rightarrow {H}_0$,
${\rho}\rightarrow \infty$ and $|p|\rightarrow \infty$. And, when $\bar{H}_0<0$, one gets
 ${H}\rightarrow {H}_0$ (contracting phase),
${\rho}\rightarrow -\infty$ and $|P|\rightarrow \infty$.


In order to qualitatively study the system it is
quite convenient, as in \cite{aw86,w85}, to
perform the variable change $\bar{p}\equiv\sqrt{|\bar{H}|}$.
After what, the semiclassical Friedmann and conservation equations become
\begin{eqnarray}\label{a19}
\bar{p}''=
-\partial_{\bar{p}}W(\bar{p},\bar{\rho})
-3\epsilon
p^2(\bar{p}^{'})^2,\quad {\bar{\rho}}'=\bar{H}\bar{\phi}_0^2\frac{\bar{\varphi}^2}{\bar{\psi}^2},
\end{eqnarray}
where
${W}(\bar{p},\bar{\rho})=\frac{\beta}{8\alpha}\left(\bar{p}^2(1-\frac{1}{3}\bar{p}^4)
+\frac{\bar{\rho}}{\bar{p}^2}\right)$, and $\epsilon\equiv$ sign $(\bar{H})$.

For positive values of $\bar{\rho}$,
the potential ${W}$ (Fig.~3 of Ref.~\cite{w85}), has a unique zero, at
$\bar{p}_0=\left(3/2\right)^{1/4}\left(1+\sqrt{1+\frac{4}{3}\bar{\rho}}\right)^{1/4}$,
and two critical points, at
$\bar{p}_{\pm}=\left(\frac{1\pm\sqrt{1-{4}\bar{\rho}}}{2}\right)^{1/4}$
($\bar{p}_-<\bar{p}_+$). Thus, for $\bar{\rho}>1/4$ there are no
critical points, being the potential strictly increasing, from
$-\infty$ to $\infty$. For $\bar{\rho}<1/4$, the potential satisfies
$W(0)=-\infty$ and ${W}(\infty)=\infty$, and exhibits a relative maximum,
at $\bar{p}_-$, and a relative minimum, at $\bar{p}_+$ (a hollow one).
For very small values of $\bar{\rho}$, at $\bar{p}_-$ one has
$\bar{H}^2\cong\bar{\rho}$, that is, the system is close to the
Friedmann phase and, at $\bar{p}_+$, one has $|\bar{H}|\cong 1$,
that is, the system is close to the de Sitter phase.
On the other hand, for negative values of $\bar{\rho}$, the potential only has a critical point
at $\bar{p}_+$, and satisfies $W(0)=W(\infty)=\infty$.

Now, assume that, initially, the system has an energy density which is positive, and that it is in the expanding phase (what does happen nowadays). Then,
since in the expanding phase the energy density increases, this means that the slope of the potential is more steep and thus the system can
evolve to the contracting phase. When it enters that phase, the energy density decreases and even it could be negative; if so,  the system
is confined in the decreasing phase because the potential satisfies $W(0)=W(\infty)=\infty$.

What is important to stress here is that the system cannot remain all the time in the expanding phase due to the form of the potential, and also the fact that
the energy density is increasing, in this phase. Three different situations may occur:
\begin{enumerate}
 \item The system may develop a singularity in a finite time (type III singularity). This comes from Eq.~(18).
\item The system may enter in the decreasing phase and the energy density becomes negative, and then the system cannot abandon this phase.
In this situation the energy density could by $-\infty$ in a finite time.
\item The system may bounce infinitely many times (an oscillating universe).
\end{enumerate}
This is what one can say by analytically studying the system. What we will do in next section is to perform a corresponding numerical study, which will to show that
only the first two situations are actually possible.

\section{Numerical analysis}
In this section we  numerically integrate the system (\ref{a15}), assuming that initially the system is in the Friedmann phase, that is,
at time $\bar{t}=0$ the variables $(\bar{H}(0),\bar{Y}(0),\bar{\psi}(0),\bar{\varphi}(0))$ satisfy the constrains
\begin{eqnarray}\label{a20}
 \bar{H}^2(0)=\frac{1}{3}\left(-\bar{Y}(0)+\frac{\bar{V}_0}{\bar{\psi}^2(0)}\right); \qquad
\bar{Y}(0)=\frac{\bar{\phi}_0^2}{2}\frac{\bar{\varphi}^2(0)}{\bar{\psi}^2(0)}.
\end{eqnarray}
This means that the initial conditions depend on two variables. Next, to perform our calculations we choose as variables $(\bar{\psi}(0),\bar{\varphi}(0))$, and also the following values for the parameters: $\bar{V}_0=24$ and $\bar{\phi}_0=4/\sqrt{3}$.
Note that, from Eq.~$(\ref{a20})$, with our choose of parameters the variable $\bar{\varphi}(0)$ belongs to the interval $[-3,3]$ and $\bar{\psi}(0)$ belongs to $(0,\infty)$.

In Fig.~$2$ we plot three simulations for different values of $\beta/\alpha$ (the system (\ref{a15}) depends on this quotient), the first being for
$\beta/\alpha=-0.5$, the second  for $\beta/\alpha=-1$,
and the last one for $\beta/\alpha=-10$. The blue color means initial conditions which drive  to a singularity
of the form given by Eqs.~(\ref{a17}) and  (\ref{a18}) in the decreasing phase, that is, the Hubble parameter is negative and the energy density
diverges to minus infinity. On the other hand, the red color means initial conditions which drive to a type III singularity.

In Fig.~$3$ we have integrated the system (\ref{a15}) for $\beta/\alpha=-10$, and we show the evolution of the Hubble parameter and of the
energy density. In the first two plots the initial conditions
are taken in the blue region  of Fig.~$1$, given a universe evolving,
at late times, in the decreasing phase with an energy density which diverges at late times. The last two plots correspond to initial conditions taken in the red region  of Fig.~$1$, and they show a type III singularity.

\begin{figure}[h]
\begin{center}
\includegraphics[scale=0.30]{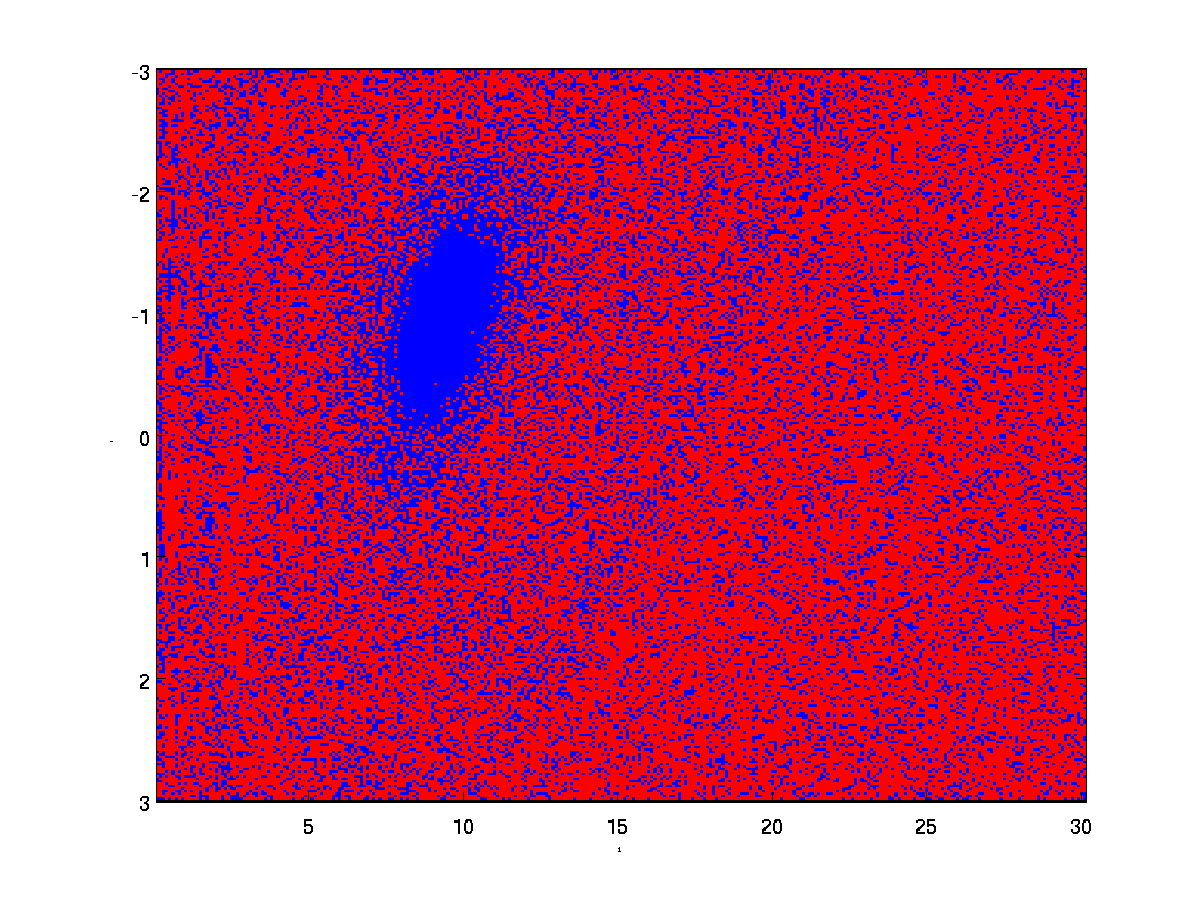}
\includegraphics[scale=0.30]{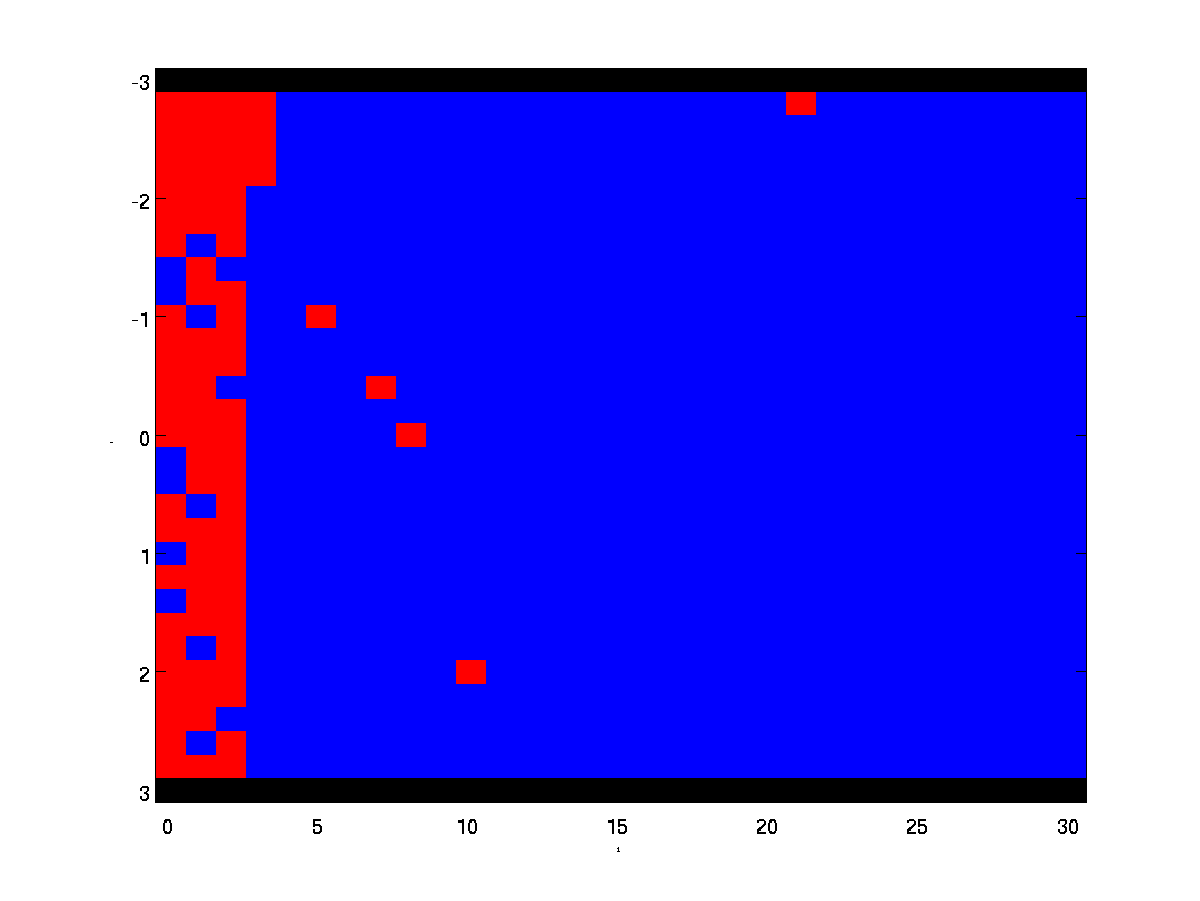}
\includegraphics[scale=0.30]{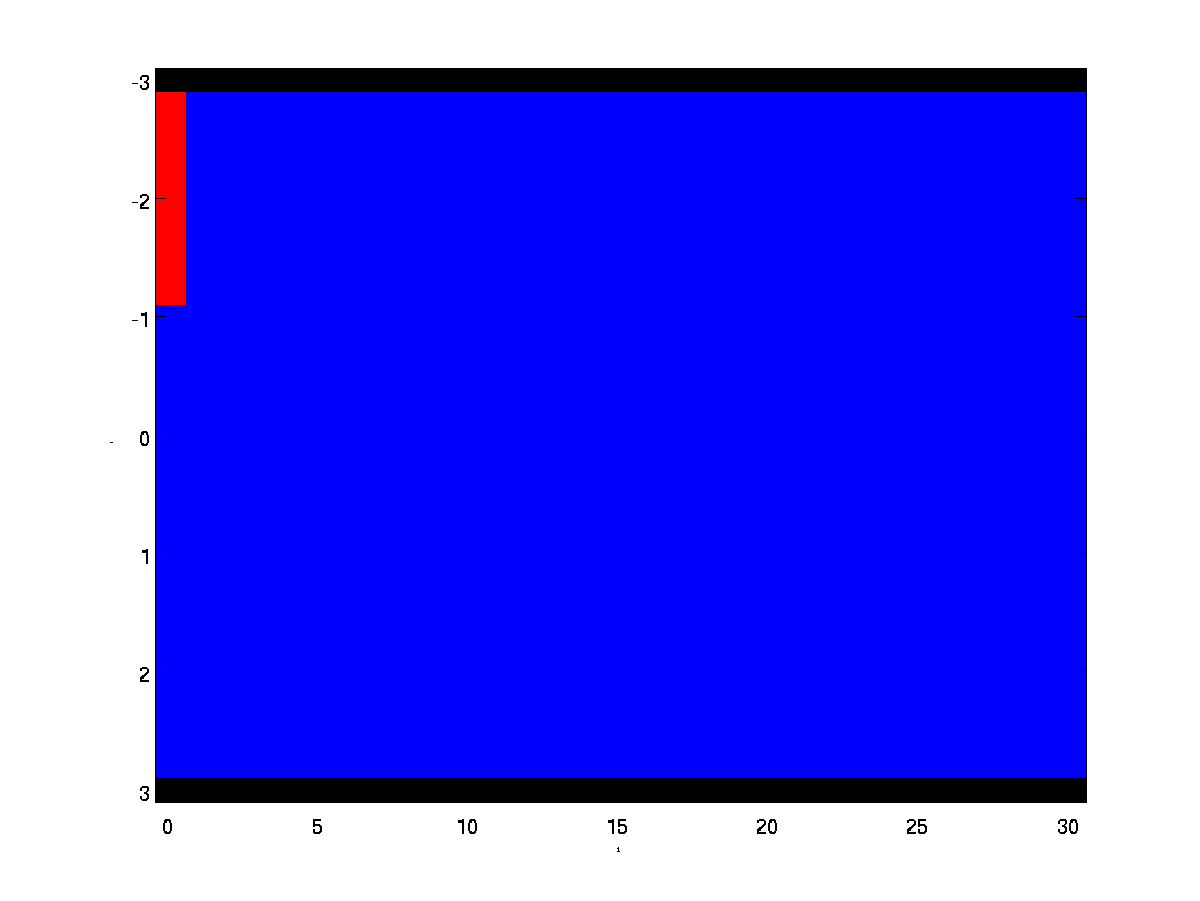}
\end{center}

\vskip-6mm
\caption{{Three different simulations, for the values $\beta/\alpha=-0.5$,  $\beta/\alpha=-1$,
and  $\beta/\alpha=-10$, respectively. Red points mean initial conditions driving to a type III singularity. Blue points, initial conditions
driving to a singularity in the decreasing phase.}}
\end{figure}

\vspace{1cm}

\begin{figure}[htb]
\begin{center}
\includegraphics[scale=0.30]{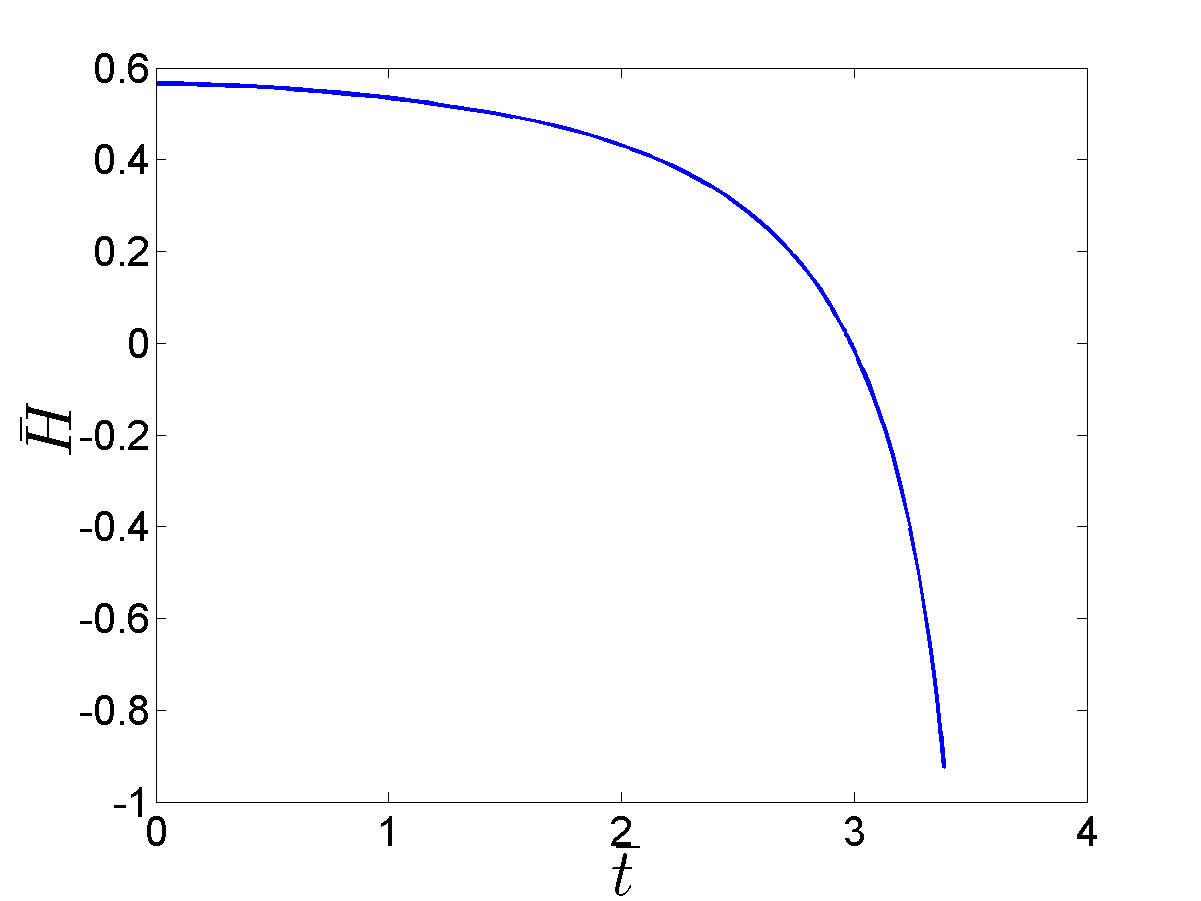}
\includegraphics[scale=0.30]{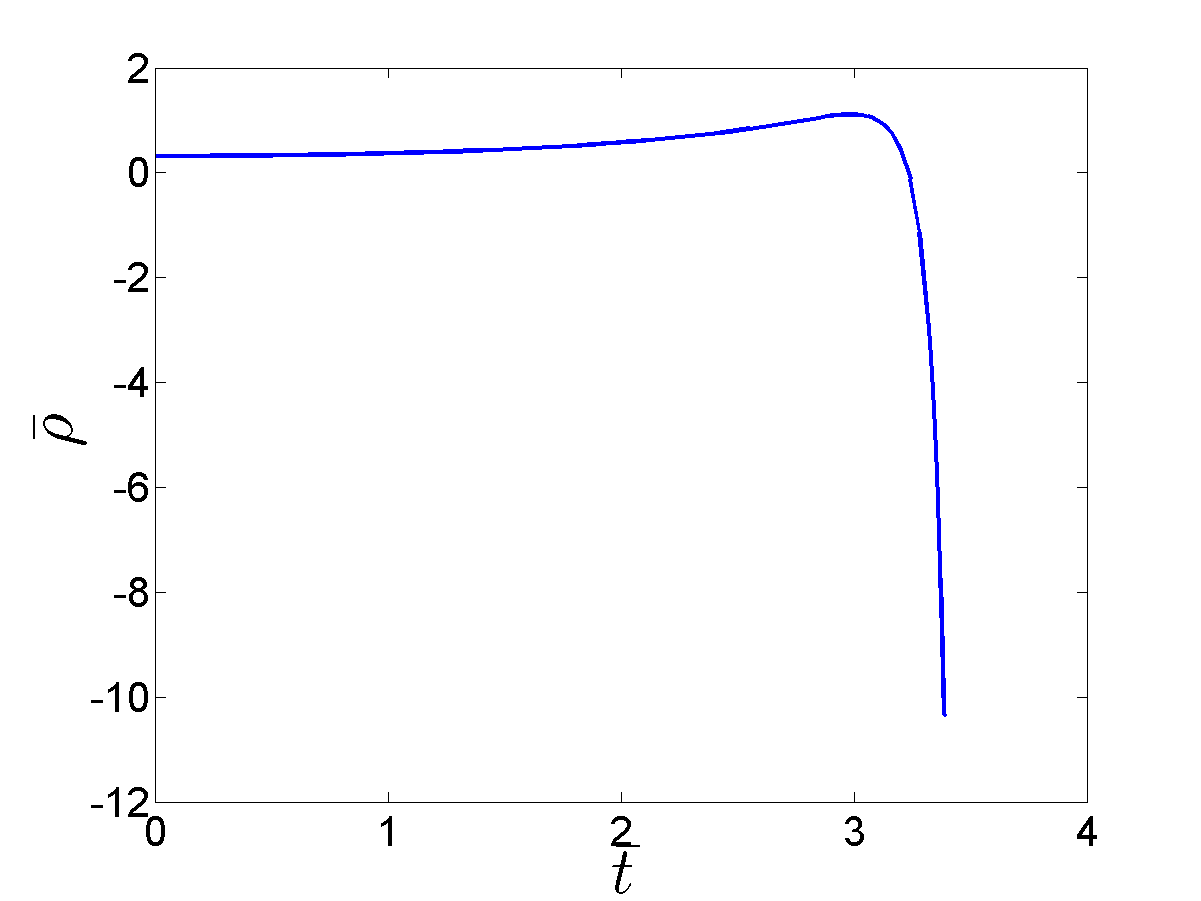}
\includegraphics[scale=0.30]{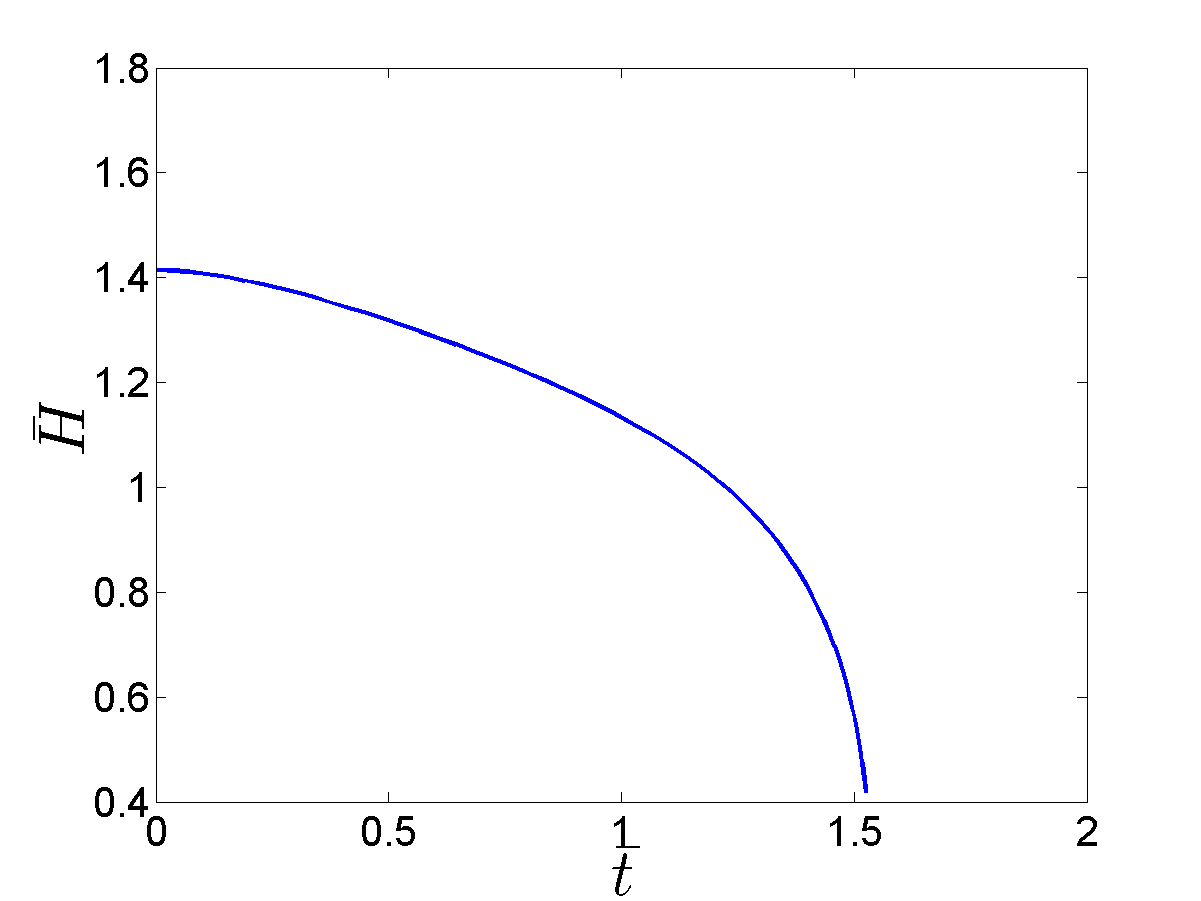}
\includegraphics[scale=0.30]{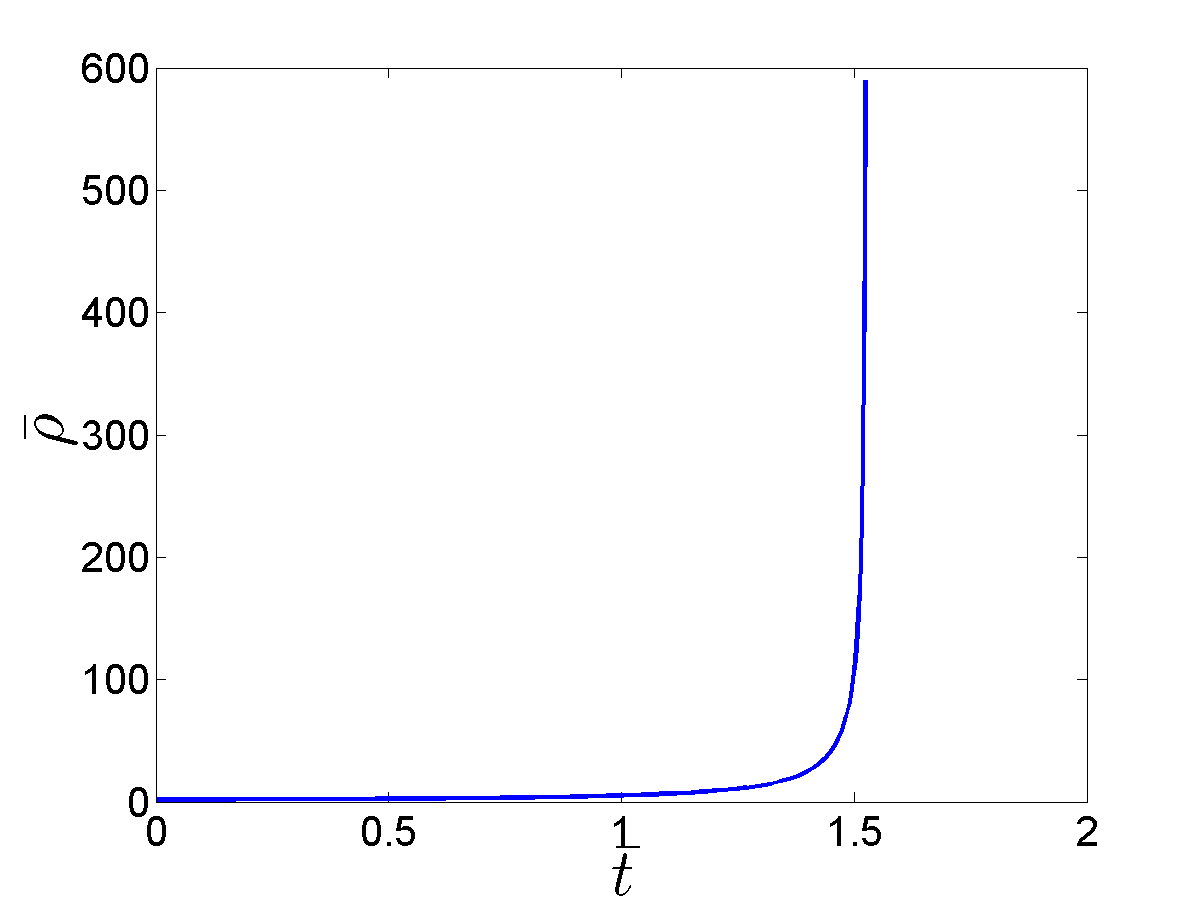}
\end{center}

\vskip-6mm
\caption{{Plots of $\bar{H}$ and $\bar{\rho}$ for $\beta/\alpha=-1$. The first two have the same initial condition, corresponding to a blue point in Fig.~$1$. The last two plots have the same initial condition, that corresponds to a red point in Fig.~$1$. }}
\end{figure}

\section{Phantom fields in Loop Quantum  Cosmology}
For the flat FRW spacetime, Einstein's theory is obtained from the Lagrangian
${\mathcal L}=\frac{1}{2}Ra^3+{\mathcal L}_{matter}$,  where $a$ denotes the scalar factor and
${\mathcal L}_{matter}=a^3P=a^3\left(-\frac{1}{2}\dot{\phi}^2-V(\phi)\right)$.
This Lagrangian can be written as follows ${\mathcal L}=3\left(\frac{d{(\dot{a} a^2)}}{dt}-\dot{a}^2 a\right)+a^3P$, what means that the same theory
is obtained avoiding the total derivative, which gives the Lagrangian ${\mathcal L}_{E}=-3\dot{a}^2 a+a^3P$.
The conjugate momentum of the scale factor is  then given by
$p=\frac{\partial {\mathcal L}_{E}}{\partial\dot{a}}=-6\dot{a}a$, and thus the Hamiltonian is
\begin{eqnarray}\label{A}
{\mathcal H}_{E}=
\dot{a}p+a^3\frac{\partial P}{\partial\dot{\phi}}\dot{\phi}- {\mathcal L}_{E}=-\frac{p^2}{12 a}+a^3\rho=-3H^2a^3+a^3\rho.\end{eqnarray}

On the other hand, in loop cosmology the following effective Hamiltonian, which captures the underlying loop quantum dynamics, is considered
\cite{as11,s09,s09a}
\begin{eqnarray}\label{B}
{\mathcal H}_{LQC}=-3{\mathcal V}\frac{\sin^2( \lambda \beta)}{\gamma^2\lambda^2}+ {\mathcal V}\rho ,
\end{eqnarray}
where $\gamma$ is the Barbero-Immirzi parameter  and  $\lambda$ is a
parameter with dimensions of length, which is determined by invoking
the quantum nature of the geometry, that is, through identification of its square with the
minimum eigenvalue of the area operator in LQG, which gives as a result $\lambda\equiv
\sqrt{\frac{\sqrt{3}}{4}\gamma}$ (see \cite{s09a}).
Here $V$ is the physical
volume ${\mathcal V}=a^3$ and $\beta$ is canonically conjugated to ${\mathcal V}$, and satisfies $\{\beta,{\mathcal V}\}=\frac{\gamma}{2}$,
where $\{,\}$ is the Poisson bracket.

The Hamiltonian constraint is then given by
$\frac{\sin^2( \lambda\beta)}{\gamma^2\lambda^2}=\frac{\rho}{3}$, and the  Hamiltonian equation yields the identity:
\begin{eqnarray}\label{B1}
\dot{{\mathcal V}}=\{{\mathcal V},{\mathcal H}_{LQC}\}=-\frac{\gamma}{2}\frac{\partial{\mathcal H}_{LQC}}{\partial\beta}\Longleftrightarrow
H= \frac{\sin(2\lambda \beta)}{2\gamma\lambda}\Longleftrightarrow \beta=
\frac{1}{2\lambda}\arcsin(2\lambda\gamma H).
\end{eqnarray}
Writing this last equation as  $H^2=\frac{\sin^2(\lambda \beta)}{\gamma^2\lambda^2}(1-\sin^2( \lambda \beta))$, and
using the Hamiltonian constraint ${\mathcal H}_{LQC}=0\Longleftrightarrow \frac{\sin^2( \lambda \beta)}{\gamma^2\lambda^2}=\frac{\rho}{3}$,
one gets the following modified Friedmann equation in loop quantum cosmology
\begin{eqnarray}\label{B2}
H^2=\frac{\rho}{3}\left(1-\frac{\rho}{\rho_c}\right)
\Longleftrightarrow \frac{H^2}{\rho_c/12}+\frac{(\rho-\frac{\rho_c}{2})^2}{\rho_c^2/4}=1,
\end{eqnarray}
being $\rho_c\equiv \frac{3}{\gamma^2\lambda^2}$.
This equation, together with the conservation equation $-\ddot{\phi}-3H\dot{\phi}+\frac{d V}{d{\phi}}=0 $, determine the dynamics of the universe in loop cosmology.

From the equation of the ellipse, one can easily check that the Hubble parameter belongs to the interval $[-\rho_c/12,\rho_c/12]$, and the energy density,  $\rho$, to
$[0,\rho_c]$, what means that there is not Big Rip. In fact, an exhaustive study of the potential $V=V_0e^{-2\phi/\phi_0}$ was performed in \cite{nw07,sg07}.

But here a problem appears. It is well-known that the  current cosmological theories are built from two invariants, the scalar curvature $R=6\left(\dot{H}+2H^2\right)$
and the Gauss-Bonnet curvature invariant
$G=24H^2\left(\dot{H}+H^2\right)$. For example,  in Gauss-Bonnet gravity \cite{cenoz06}  the Lagrangian ${\mathcal L}_{MG}=a^3f(R,G)+a^3P$ is used,
 and semiclassical gravity, when one takes into account the quantum effects due a massless conformally coupled field (see for instance \cite{hae11}),
is based in the trace anomaly $T_{vac}= \alpha\Box  R-\frac{\beta}{2}G$ (being $\alpha>0$ and $\beta<0$ two renormalization coefficients). However,
from the Legendre transformation
\begin{eqnarray}\label{B3}{\mathcal H}_{LQC}=-\frac{2}{\gamma}\dot{\mathcal V}\beta
+{\mathcal V}\frac{\partial P}{\partial\dot{\phi}}\dot{\phi}-{\mathcal L}_{LQC}\end{eqnarray}
one gets, in terms of the standard variables, the following Lagrangian
\begin{eqnarray}\label{B4}
 {\mathcal L}_{LQC}=-\frac{3a^3H}{\gamma\lambda}\arcsin(2\lambda\gamma H)+\frac{3a^3}{2\gamma^2\lambda^2}\left(1-\sqrt{1-4\gamma^2\lambda^2H^2}
\right)+a^3 P,
\end{eqnarray}
which  is not invariant. This is in disagreement with one of the main principles of general relativity.

From these observations, one can conclude that
the modified Friedmann equation (\ref{B2}) does not stand in this form, because it has been obtained assuming that (\ref{B}) is the Hamiltonian of the
system which is in contradiction with the invariance of General Relativity. Finally, we also conclude that the results obtained from this modified Friedmann equation need a deep revision (for more details, see \cite{h12}).

\section{Discussion and comparison with the phantom fluid model}
In \cite{hae11} we have studied in detail the case of a phantom fluid modeled by the EoS $P=\omega \rho$ with $\omega<-1$. There, we have shown that,
in the case $-1\leq\frac{\beta}{3\alpha}<0$,
there exists a one parameter family of solutions which evolves into the contracting Friedmann phase at late times
and only a particular solution asymptotically converging towards the de contracting de Sitter universe. All the other solutions enter into the contracting phase and become singular
at finite time, satisfying $\lim_{t\rightarrow t_s}H(t)=-\infty$ and $\lim_{t\rightarrow t_s}\rho(t)=0$.
On the other hand, we have shown in \cite{hae11}, that for $-1>\frac{\beta}{3\alpha}$ almost
all solutions describe a universe bouncing infinitely many times (an oscillating
universe).

In the present paper, by studying a phantom field we have shown, both analytically and numerically,  that all solutions are singular.
Some of them display Type III singularities and the other ones are singular in the contracting phase, satisfying  ${H}(t)\rightarrow {H}(t_s)<0$,
${\rho}(t)\rightarrow -\infty$ and $|P(t)|\rightarrow \infty$,  when $t\rightarrow t_s$.

The difference comes from the fact that, for a phantom fluid, when one considers the dynamics in $\R^3$ using the coordinates  $(H,\dot{H},\rho)$,
 the manifold $\rho=0$ is invariant. More precisely, the half plane $\rho=0$ with $H>0$  is a repeller, whereas when
$H>0$ it is an attractor. This means that, at late time, all the solutions go towards this half plane. Moreover, in the
contracting phase there is a critical point $(-\sqrt{-1/\beta},0,0)$
(the contracting de Sitter universe)
which restricted to the plane $\rho=0$ is a repeller. This means that only a solution tends asymptotically towards this point, while all
the other escape towards infinity in finite time
(this was proven in \cite{hae11}). The same does not happen for a phantom field where the manifold $\rho=0$ is not invariant, and eventually, the system crosses this manifold, i.e.,
it can have negative energy density, and then
it cannot leave the decreasing phase, becoming singular at finite time, as we have shown, again numerically and analytically.

\vspace {1cm}

\noindent {\bf Acknowledgments.}
This investigation has been
supported in part by MICINN (Spain), projects MTM2011-27739-C04-01,
MTM2009-14163-C02-02, and FIS2010-15640, by the CPAN Consolider
Ingenio Project, and by AGAUR (Generalitat de Ca\-ta\-lu\-nya),
contracts 2009SGR 345, 994 and 1284. EE was also supported by MICINN 
(Spain), contract PR2011-0128, and his research was partly carried out while on 
leave at the Department of Physics and Astronomy, Dartmouth College, 6127 Wilder 
Laboratory, Hanover, NH 03755, USA.

\end{document}